\documentclass[12pt,preprint]{aastex}

\usepackage{graphicx}
\usepackage{txfonts}
\usepackage{longtable}
\usepackage{lscape}
\usepackage{natbib}

\shorttitle{The scattering polarization of H~{\sc i} and He~{\sc ii} 
Ly$\alpha$ lines}
\shortauthors{Belluzzi, Trujillo Bueno, \v{S}t\v{e}p\'an}

\begin{document}

\title{The scattering polarization of the Ly$\alpha$ lines of H~{\sc i} and 
He~{\sc ii} \\
taking into account PRD and $J$-state interference effects}

\author{{\sc Luca Belluzzi}\altaffilmark{1,2}, 
{\sc Javier Trujillo Bueno}\altaffilmark{1,2,3}, 
{\sc and Ji\v{r}\'i \v{S}t\v{e}p\'an}\altaffilmark{4}}
\altaffiltext{1}{Instituto de Astrof\'isica de Canarias, E-38205 La Laguna, 
Tenerife, Spain}
\altaffiltext{2}{Departamento de Astrof\'isica, Facultad de F\'isica, 
Universidad de La Laguna, Tenerife, Spain}
\altaffiltext{3}{Consejo Superior de Investigaciones Cient\'ificas, Spain}
\altaffiltext{4}{Astronomical Institute ASCR, Ond\v{r}ejov, Czech Republic}

\begin{abstract}
Recent theoretical investigations have pointed out that the cores of the 
Ly$\alpha$ lines of H~{\sc i} and He~{\sc ii} should show measurable scattering 
polarization signals when observing the solar disk, and that the magnetic 
sensitivity, through the Hanle effect, of such linear polarization signals 
is suitable for exploring the magnetism of the solar transition region. 
Such investigations were carried out in the limit of complete frequency 
redistribution (CRD) and neglecting quantum interference between the two upper 
$J$-levels of each line. 
Here we relax both approximations and show that the joint action of partial 
frequency redistribution (PRD) and $J$-state interference produces much more 
complex fractional linear polarization ($Q/I$) profiles, with large amplitudes 
in their wings.
Such wing polarization signals turn out to be very sensitive to the 
temperature structure of the atmospheric model, so that they can be exploited 
for constraining the thermal properties of the solar chromosphere.
Finally, we show that the approximation of CRD without $J$-state interference 
is however suitable for estimating the amplitude of the linear polarization 
signals in the core of the lines, where the Hanle effect operates.
\end{abstract}


\section{Introduction}
In the transition region between the chromosphere and corona of the Sun, the 
kinetic temperature suddenly jumps from $10^4$~K to $10^6$~K and the plasma 
changes from partially to practically fully ionized.
Although various physical mechanisms have been proposed for explaining 
this temperature increase, the lack of reliable measurements of key physical 
quantities, such as the magnetic field, represents today the most serious 
limitation to a better comprehension of the role played by this boundary 
region on the heating of the solar corona.
In order to make measurements of the quantities that remain basically unknown,
we need (1) to identify observables sensitive to the physical conditions of 
the solar transition region, (2) to develop the instruments needed for 
measuring such observables, and (3) to infer the relevant physical quantities 
(e.g., the strength and orientation of the magnetic field) through realistic 
modeling of the measured observables. 

Recently, \citet{JTB11} argued that the hydrogen Ly$\alpha$ line at 1216~\AA\ 
should show measurable scattering polarization signals when observing the solar 
disk, and that via the Hanle effect the line-center amplitudes of the $Q/I$ and 
$U/I$ linear polarization profiles must be sensitive to the strength and 
orientation of the magnetic field in the solar transition region (with good 
sensitivity to magnetic field strengths between 10 and 100~G). 
In a subsequent paper, \citet{JTB12} pointed out that significant line-center 
scattering polarization signals are to be expected also for the Ly$\alpha$ line 
of He~{\sc ii} at 304~\AA, and that for the magnetic field strengths expected 
at transition region heights (i.e., $B{\lesssim}100$~G outside active regions) 
the He~{\sc ii} 304~\AA\ line, due to its very large Einstein coefficient, is 
immune to the Hanle effect (i.e., it is a line whose observed linear 
polarization pattern could be used as a reference for facilitating the 
identification of the observational signature of the Hanle effect in the 
hydrogen Ly$\alpha$ line). 
The possibility that scattering processes produce measurable linear 
polarization signals in these and other emission lines of the solar transition 
region is of great scientific interest, because it is mainly through the Hanle 
effect that we may hope to explore the magnetism of the upper chromosphere and 
transition region of the Sun. 
As a matter of fact, with few exceptions, the polarization signals induced by 
the Zeeman effect in transition region lines are expected to be very weak 
(e.g., in typical semi-empirical models of the solar atmosphere a 
volume-filling magnetic field of 100~G inclined by ${45}^{\circ}$ with respect 
to the line of sight (LOS) produces in both Ly$\alpha$ lines circular 
polarization $V/I$ signals significantly smaller than 0.1\%, while the 
contribution of the transverse Zeeman effect to their linear polarization is 
insignificant).

The conclusions of \citet{JTB11,JTB12} were obtained through detailed radiative 
transfer (RT) calculations in semi-empirical and hydrodynamical models of the 
solar atmosphere, applying the quantum theory of polarization described in the 
monograph by \citet{Lan04}. 
This density-matrix theory is based on the approximation of complete frequency 
redistribution (CRD), which neglects correlations between the frequencies of 
the incoming and outgoing photons in the scattering events. 
In addition, \citet{JTB11,JTB12} neglected quantum interference between the 
$^2{\rm P}_{1/2}$ and $^2{\rm P}_{3/2}$ upper levels of such Ly$\alpha$ lines. 
Both approximations should be suitable for estimating the linear polarization 
amplitudes at the line center, which is where the Hanle effect operates. 
However, in strong resonance lines the joint action of partial frequency 
redistribution (PRD) and $J$-state interference can produce important 
observational signatures in the wings of the fractional linear polarization 
($Q/I$) profiles \citep[e.g.,][]{Smi11a,Smi11b,Bel12}.

The aim of the present paper is to investigate the impact of PRD and $J$-state 
interference effects on the scattering polarization of the Ly$\alpha$ lines of 
H~{\sc i} and He~{\sc ii}. 
To this end, we apply the same theoretical framework and RT code that 
\citet{Bel12} developed for investigating the scattering polarization pattern 
across the $h$ and $k$ lines of Mg~{\sc ii}.
Like the Mg~{\sc ii} $h$ and $k$ lines, also the above-mentioned 
Ly$\alpha$ lines result from two transitions between a common lower level with 
angular momentum $J=1/2$ and two upper levels with $J=1/2$ and $J=3/2$.
However, in contrast with the case of the Mg~{\sc ii} resonance lines, the two 
components of each Ly$\alpha$ line are fully blended since the fine-structure 
splitting between the two upper $J$-levels is much smaller than the 
Doppler line width.\footnote{Note that the fine-structure splitting 
between the two upper $J$-levels, though small, is in any case much 
larger than the levels natural width.}

\section{Formulation of the problem}
This investigation has been carried out within the same theoretical framework   
that has been recently developed by \citet{Bel12} for investigating the 
polarization properties of the Mg~{\sc ii} $h$ and $k$ lines.
The reader is referred to that paper for more details on the theoretical tools 
that we have applied.

We consider a {\em two-term} atomic model both for H~{\sc i} and He~{\sc ii}, 
the lower term being composed of the ground level ($^2{\rm S}_{1/2}$) and the 
upper term of the upper levels ($^2{\rm P}_{1/2}$ and $^2{\rm P}_{3/2}$).
Such atomic model allows us to take into account the fine structure of the 
Ly${\alpha}$ line, as well as the contribution of the $J$-state interference 
between the two upper $J$-levels.

The Non-LTE RT problem is formulated within the redistribution matrix 
formalism. We consider a linear combination of the $R_{\rm II}$ and 
$R_{\rm III}$ redistribution matrices (which describe purely coherent 
scattering and completely redistributed scattering in the atom rest frame, 
respectively), for a two-term atom with unpolarized and infinitely sharp 
lower term, in the absence of magnetic fields.
The redistribution matrix $R_{\rm II}$ in the observer frame is calculated
starting from the expression derived by \citet{Lan97} in the atom rest frame, 
and taking Doppler redistribution into account \citep[see Eq.~(3) of][]{Bel12}.
The redistribution matrix $R_{\rm III}$ is calculated in the limit of complete 
frequency redistribution (CRD) in the observer frame, starting from the CRD 
theory for a two-term atom presented in \citet{Lan04}.
The effect of inelastic and superelastic collisions is included under the same 
approximations made in \citet{Bel12}. The collisional rates for the H~{\sc i} 
Ly${\alpha}$ line have been calculated form \citet{Prz04}, those for the 
He~{\sc ii} Ly${\alpha}$ line from \citet{Jan87}.  

We consider the branching ratios for $R_{\rm II}$ and $R_{\rm III}$ derived 
for the two-level atom case \citep[see][]{Bom97a,Bom97b}.
Neglecting depolarizing collisions, such branching ratios are given by $\alpha$ 
and $(1-\alpha)$, respectively, with\footnote{Note that the branching ratios 
$\alpha$ and $(1-\alpha)$, defined according to Eq.~(1), do not include the 
factor $(1-\epsilon)$, with $\epsilon$ the photon destruction probability.
This factor is directly included in the redistribution matrices.} 
\begin{equation}
	\alpha=\frac{\Gamma_R + \Gamma_I}{\Gamma_R + \Gamma_I + \Gamma_E} \; .
\end{equation}
The quantities $\Gamma_R$, $\Gamma_I$, and $\Gamma_E$ are the broadening widths 
of the upper level due to radiative decays, superelastic collisions, and 
elastic collisions, respectively.\footnote{Our theoretical formulation is based 
on the hypothesis that such broadening withs are identical for the two levels 
of the upper term. This is a very good approximation for the model atoms 
considered in this work.}

The variation of $\alpha$ as a function of height for the two lines under 
investigation, in the semi-empirical solar atmosphere model C of \citet{Fon93}
(hereafter, FAL-C), is shown in the left panel of Fig.~1. 
As pointed out in \citet{JTB12}, and as it can be seen from the right panel of 
Fig.~1, the He~{\sc ii} Ly$\alpha$ line forms in a thin layer of the transition 
region (approximately, between about 2000 and 2200~km). 
As shown in the left panel of Fig.~1, at these heights $\alpha$ is 
essentially equal to unity so that the assumption of purely coherent scattering 
in the atom rest frame is an excellent approximation for the modeling of this 
line.
The situation is slightly different as far as the H~{\sc i} Ly${\alpha}$ 
line is concerned. 
While the core of this line forms in the transition region, just a few 
kilometers below the core of the He~{\sc ii} Ly$\alpha$ line, its broad 
wings form deeper in the chromosphere, where $\alpha$ assumes values 
appreciably smaller than one, though larger than 0.9. 
The approximation of purely coherent scattering in the atom rest frame is 
thus not so good for this line, and its modeling requires to take both 
$R_{\rm II}$ and $R_{\rm III}$ into account.

The polarization profiles of the two lines under investigation are calculated 
by applying a Non-LTE RT code based on the angle-averaged expressions of 
$R_{\rm II}$ and $R_{\rm III}$, and including the contribution of an 
unpolarized continuum.
The numerical method of solution is a direct generalization to the PRD case of 
the Jacobian iterative scheme presented in \citet{JTB99}.
The initialization of the iterative calculation is done using the 
self-consistent solution of the corresponding unpolarized problem, obtained 
by applying Uitenbroek's (2001) RT code.
The same code is used to calculate the continuum opacity (including the UV line 
haze) and emissivity. 

We point out that due to the negligible impact of $J$-state interference 
on Stokes-$I$, the PRD intensity $I(\lambda)$ profiles obtained with our code 
coincide with those computed using Uitenbroek's (2001) code (which neglects 
$J$-state interference).
As we shall see below, $J$-state interference effects, together with those 
caused by PRD, have however an important impact on the $Q/I$ profiles.

\section{The scattering polarization profile of the H~{\sc i} Ly$\alpha$ 
line at 1216~\AA}
Figure~2 shows the PRD $Q/I$ profiles of the H~{\sc i} Ly$\alpha$ line,
obtained by taking into account and neglecting quantum interference between 
the two upper $J$-levels. 
The calculations have been performed in the FAL-C atmospheric model, for a 
line-of-sight (LOS) with $\mu \equiv \cos\theta = 0.3$ (with $\theta$ the 
heliocentric angle).

The two profiles perfectly coincide in the core of the line where $J$-state 
interference does not produce any observable signature \citep[cf.,][]{Bel11}.
In this spectral region, our PRD profile shows a good agreement with the 
corresponding CRD profile calculated by \citet{JTB11} (see the left panel 
of their Fig.~2).\footnote{We point out that our PRD calculations and the CRD 
calculations of \citet{JTB11} were made fixing the total neutral hydrogen 
number density to the values tabulated in the FAL-C model.}

Outside the line core, PRD effects produce a complex linear polarization 
profile with extended wings, while $J$-state interference plays a  
significant role, producing much larger polarization amplitudes with respect 
to the case in which it is neglected. 
The profile obtained by taking $J$-state interference into account shows in 
particular two narrow negative peaks at approximately $\pm 0.4$~{\AA} from 
line center, with an amplitude of about -7\%, and two broad negative lobes 
with a minimum of about -6\% at approximately $\pm 10$~{\AA} from line center. 
As seen in Fig.~3, the shape of the $Q/I$ profile remains qualitatively the 
same for line of sights corresponding to larger $\mu$ values, although the 
$Q/I$ amplitudes are clearly smaller.

Figure~4 shows the sensitivity of the scattering polarization profile of the 
hydrogen Ly$\alpha$ line to the thermal structure of the model atmosphere.
We compare the results for the atmospheric models C, F and P of \citet{Fon93}, 
which can be considered as illustrative of quiet, network and plage regions. 
Such sensitivity is due to the fact that the anisotropy of the incident 
radiation field (which induces atomic level polarization in the 
$^2{\rm P}_{3/2}$ level and quantum interference between its sublevels and 
those of the $^2{\rm P}_{1/2}$ level) depends on the gradient of the 
Stokes-$I$ component of the source function \citep[e.g.,][]{JTB01,Lan04}. 
Notice that especially the wing $Q/I$ signals, which are insensitive to the 
magnetic field (since the Hanle effect operates only in the line core), are 
very sensitive to the temperature structure of the model atmosphere. 
Therefore, in addition to the $I(\lambda)$ profile itself, the wings of the 
$Q/I$ profile could help us to constrain the thermal properties of the observed 
atmospheric region, and this would in turn facilitate the modeling of the 
line-core signals and the inference of the magnetic field.

\section{The scattering polarization profile of the He~{\sc ii} Ly$\alpha$ line 
at 304~\AA}
The left panel of Fig.~5 shows our PRD results for the $Q/I$ profile of the 
Ly$\alpha$ line of He~{\sc ii} calculated in the FAL-C atmospheric model, 
taking into account (solid line) and neglecting (dashed line) the impact of 
$J$-state interference. 
The line-core polarization, and its modification by the Hanle effect, can 
be safely modeled in the CRD limit, and neglecting $J$-state interference. 
However, as soon as we go outside the line-core region, the combined action of 
PRD and $J$-state interference effects produces strong $Q/I$ signals, which 
are not obtained when such physical ingredients are neglected.
The right panel of Fig.~3 shows the center to limb variation of the $Q/I$ 
profile.

Finally, in the right panel of Fig.~5 we provide information on the sensitivity 
of the $Q/I$ profile 
to the thermal structure of the atmospheric model. As can be seen, the 
wings of the $Q/I$ signals are very sensitive to the model's temperature 
structure, while the line-core signals are practically identical in the three 
atmospheric models.

\section{Understanding the impact of $J$-state interference}
The impact of $J$-state interference on the $Q/I$ profiles of various 
multiplets was first demonstrated and explained in \citet{Ste80} and considered 
in greater detail in \citet{Ste97}, Landi Degl'Innocenti \& Landolfi (2004), 
and \citet{Bel11}.

The results presented in the previous sections show that
in the wings of the lines, the amplitude of the 
$Q/I$ profiles is about a factor 3 larger when the effects of $J$-state 
interference are taken into account (see Figs.~2 and 5). 
In order to understand this behavior, we first note that far away from its 
``center of gravity'', a multiplet (in which the contribution of interference 
between different $J$-levels is taken into account) behaves in resonance 
scattering as a two-level atom transition with $J_\ell = L_\ell$ and 
$J_u = L_u$ \citep[see][]{Lan04}.
We also observe that at these frequencies scattering is purely coherent in the 
observer frame. 
Starting from the expressions of the emission coefficients given in 
\citet{Lan97}, it can be shown that, far from the center of gravity of the 
multiplet, the fractional polarization of the radiation scattered at 
$90^{\circ}$ is given by
\begin{equation}
	\left[ p_Q(\nu) \right]_{\rm int.} \equiv 
	\frac{\varepsilon_Q(\nu)}{\varepsilon_I(\nu)} = 
	\frac{3}{4} W_2(L_\ell,L_u) \, w(\nu) \; ,
\end{equation}
where the quantity $W_2(J_\ell,J_u)$ is defined in Eq.~(10.17) of 
\citet{Lan04}, $w(\nu)=\sqrt{2}J^2_0(\nu)/J^0_0(\nu)$ is the monochromatic 
anisotropy factor, and where we have neglected the contribution of $J^2_0$ to 
$\varepsilon_I$ \citep[see Eq.~(5.157) of][for the definition of the radiation 
field tensor $J^K_Q(\nu)$]{Lan04}.

From the same equations, it can be shown that for a multiplet having a single 
$J$-level in the lower term, far from its center of gravity, the fractional 
polarization obtained taking into account the contribution of the various 
lines, but neglecting interference, is given by
\begin{equation}
   \left[ p_Q(\nu) \right]_{\rm no \; int.} = \sum_{J_u} {\mathcal S}_{J_u} \, 
   \frac{3}{4} W_2(J_\ell, J_u) \, w(\nu) \; ,
\end{equation}
where ${\mathcal S}_{J_u}$ is the relative strength of the 
$J_{\ell} \rightarrow J_u$ transition (see, e.g., Eq.~(3.65) of Landi 
Degl'Innocenti \& Landolfi 2004).
For our ${^2S} - {^2P}$ multiplet, having $L_{\ell}=0$ and $L_u=1$, we then 
have
\begin{equation}
	\frac{\left[ p_Q(\nu) \right]_{\rm int.}}{\left[ p_Q(\nu) 
	\right]_{\rm no \, int.}} = 
	\frac{W_2(1,0)}{{\mathcal S}_{1/2} \, W_2(1/2,1/2) + {\mathcal S}_{3/2} 
	\, W_2(1/2,3/2)} \; .
\end{equation}
Using the numerical values $W_2(1,0)=1$, $W_2(1/2,1/2)=0$, $W_2(1/2,3/2)=1/2$,
${\mathcal S}_{1/2}=1/3$, ${\mathcal S}_{3/2}=2/3$, we easily find
$\left[ p_Q(\nu) \right]_{\rm int.} / \left[ p_Q(\nu) \right]_{\rm no \, int.} 
= 3$, in agreement with the numerical results shown in this Letter.

\section{Conclusions}
As shown in this Letter, the joint action of PRD and $J$-state interference 
effects produce complex scattering polarization $Q/I$ profiles in the 
Ly$\alpha$ lines of H~{\sc i} and He~{\sc ii}, with large polarization 
amplitudes in the wings. 
PRD effects determine the qualitative shape of the $Q/I$ profiles, while 
$J$-state interference plays a key role in producing the large polarization 
amplitudes in the wings of the lines.
Such wing polarization signals turn out to be very sensitive to the 
temperature structure of the atmospheric model. They could thus be exploited, 
in addition to the intensity profile itself, to constrain the thermal 
properties of the solar chromosphere.

As expected, the CRD approximation is only suitable for estimating the 
line-core signals. For instance, for a LOS with $\mu=0.3$ the line-center 
$Q/I$ signal of the He~{\sc ii} 304~\AA\ line calculated with PRD and 
$J$-state interference in the FAL-C atmospheric model is about $-1.3\%$, 
while it is about a factor three smaller for the hydrogen Ly$\alpha$ line. 
These line-center $Q/I$ amplitudes are in good agreement with those obtained 
by \citet{JTB12}, who assumed CRD and neglected $J$-state interference effects 
for investigating the impact of the Hanle effect on the linear polarization 
produced by scattering processes in the core of such Ly$\alpha$ lines.

Finally, it is of interest to estimate how large are the linear polarization 
amplitudes that result from the wavelength-integrated Stokes profiles 
(${\langle Q \rangle}/{\langle I \rangle}$).
It turns out that for the He~{\sc ii} line ${\langle Q \rangle}/{\langle I 
\rangle}$ is similar to the line-center amplitude (e.g., for a LOS with 
$\mu=0.3$ we have ${\langle Q \rangle}/{\langle I \rangle}{\approx}-1.3\%$ in 
the FAL-C model), while for the hydrogen Ly$\alpha$ line ${\langle Q \rangle}/
{\langle I \rangle}$ is much larger than the line-center amplitude (e.g., for a 
LOS with $\mu=0.3$ we have ${\langle Q \rangle}/{\langle I \rangle}{\approx}
-3\%$ in the FAL-C model).
Narrow band filter polarimetry (e.g., with a FWHM$\sim 1~\AA$) in these 
Ly$\alpha$ lines would provide interesting $I$, $Q/I$ and $U/I$ images over a 
large field of view. 
However, the information on the magnetic field of the solar transition region 
is encoded in the line-core polarization of the hydrogen Ly$\alpha$ line, which 
is where the Hanle effect operates. 
For this reason, the Chromospheric Ly$\alpha$ Spectropolarimeter 
\citep[see][]{Kob12} aims at doing spectropolarimetry of the hydrogen 
Ly$\alpha$ line with a spectral resolution of 0.1~\AA\ and a polarimetric 
sensitivity of 0.1\%. 
Since outside active regions the He~{\sc ii} 304~\AA\ line (which originates 
entirely within the transition region) is practically immune to the Hanle 
effect, the possibility of having in addition the information provided 
by filter polarimetry in this reference line would facilitate the diagnostics 
of the thermal and magnetic structure of the solar transition region.

\acknowledgements
Financial support by the Spanish Government through projects AYA2010-18029 
(Solar Magnetism and Astrophysical Spectropolarimetry) and CONSOLIDER INGENIO 
CSD2009-00038 (Molecular Astrophysics: The Herschel and Alma Era), and by the 
Grant Agency of the Czech Republic through grant P209/12/P741 and through 
project RVO:67985815 are gratefully acknowledged.

\begin{figure}
\centering
\includegraphics[width=\textwidth]{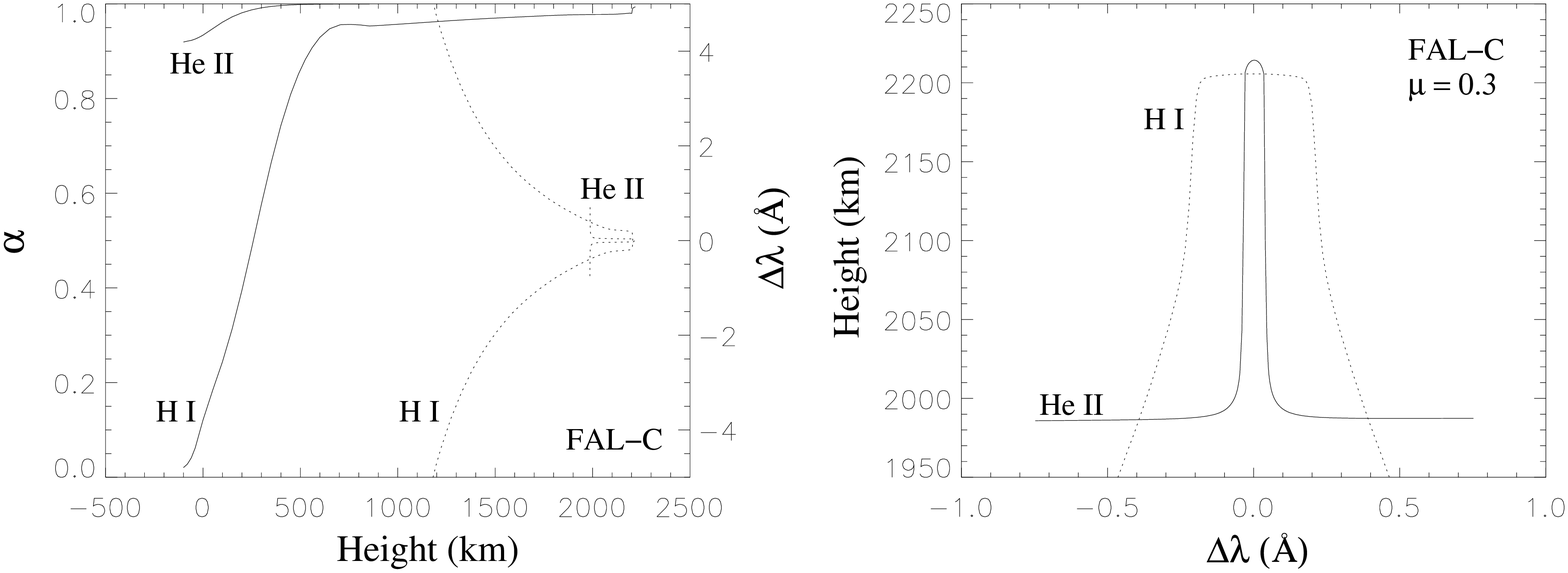}
\caption{Left panel: the solid lines show, for each Ly$\alpha$ line, the 
variation with height in the FAL-C model atmosphere of the branching ratio 
$\alpha$ of Eq.~(1).
Note that for the He~{\sc ii} 304~{\AA} line $\alpha$ is essentially equal 
to unity at chromospheric and transition region heights.
The corresponding dotted lines give the atmospheric height where, at each 
wavelength, the ensuing optical depth is unity along a LOS with $\mu=0.3$ 
(see also the right panel for a magnified visualization).}
\label{fig:figure-1}
\end{figure}

\begin{figure}
\centering
\includegraphics[width=\textwidth]{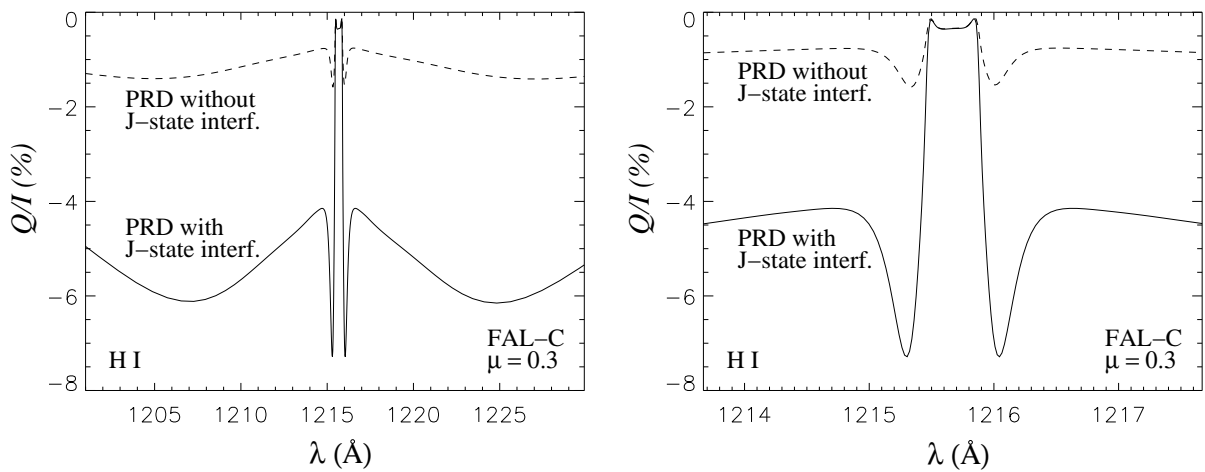}
\caption{The PRD $Q/I$ profile across the hydrogen Ly$\alpha$ line, calculated
in the FAL-C model atmosphere for a LOS with $\mu=0.3$, taking into account 
(solid line) and neglecting (dashed line) $J$-state interference.
The right panel shows in more detail the line core region.
The reference direction for positive $Q$ is the parallel to the nearest limb.} 
\label{fig:figure-2}
\end{figure}

\begin{figure}
\centering
\includegraphics[width=\textwidth]{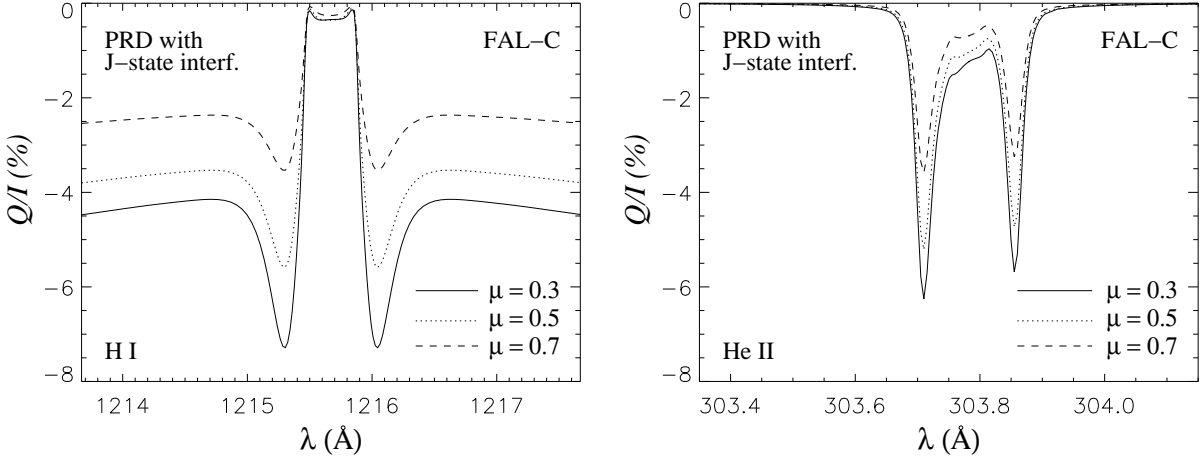}  
\caption{The center to limb variation of the $Q/I$ profile of the Ly$\alpha$ 
lines of H~{\sc I} (left panel) and He~{\sc II} (right panel), calculated in 
the FAL-C model atmosphere.
The reference direction for positive $Q$ is the parallel to the nearest limb.} 
\label{fig:figure-3}
\end{figure}

\begin{figure}
\centering
\includegraphics[width=\textwidth]{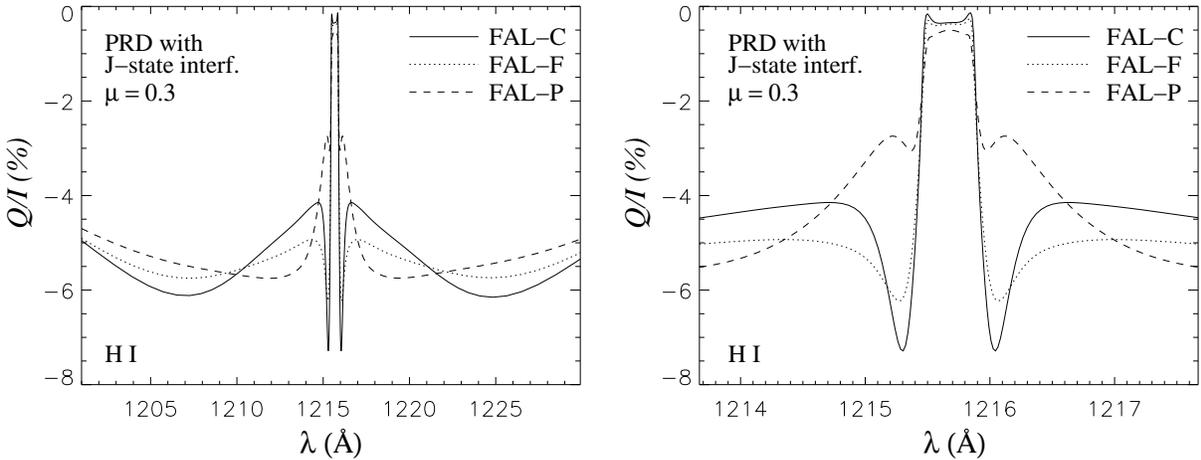}  
\caption{The $Q/I$ profile of the hydrogen Ly$\alpha$ line calculated in the 
indicated atmospheric models, taking into account PRD and $J$-state 
interference effects. 
The right panel shows in more detail the line core region.
The reference direction for positive $Q$ is the parallel to the nearest limb.}
\label{fig:figure-4}
\end{figure}

\begin{figure}
\centering
\includegraphics[width=\textwidth]{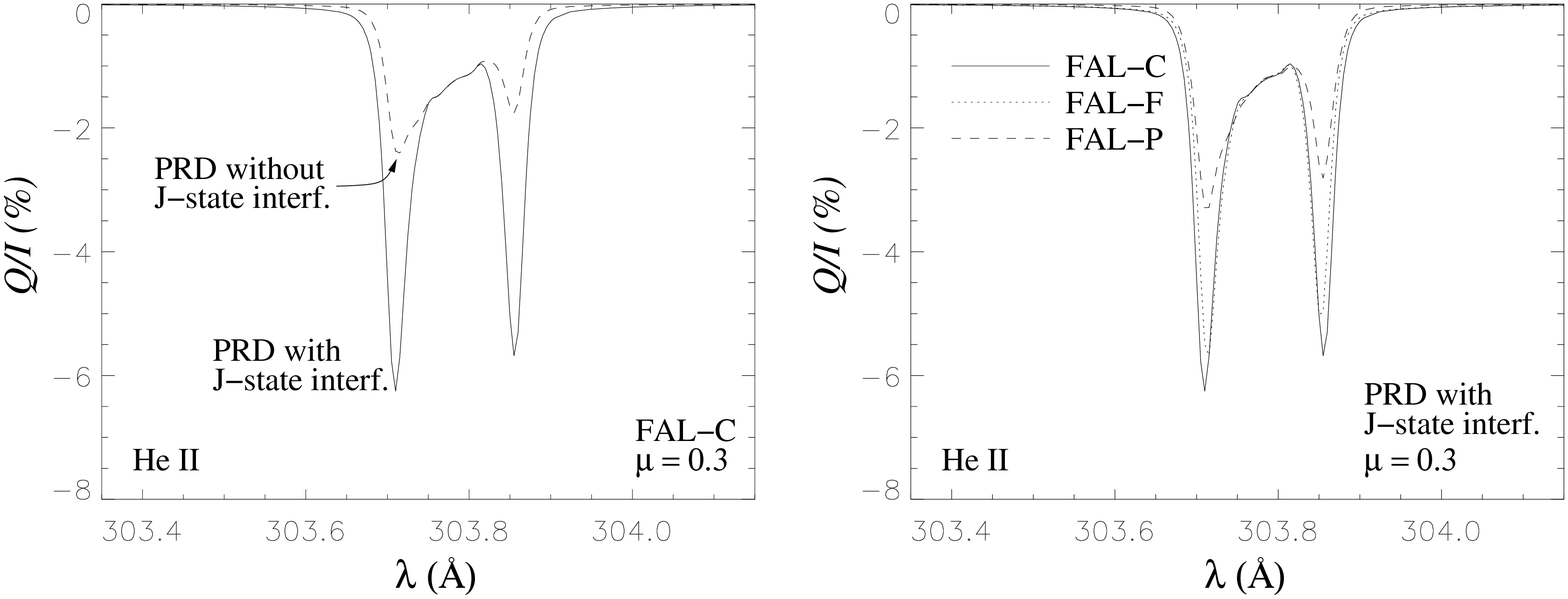}  
\caption{Left panel: the PRD $Q/I$ profile across the He~{\sc ii} Ly$\alpha$ 
line, calculated in the FAL-C model atmosphere for a LOS with $\mu=0.3$,
taking into account (solid line) and neglecting (dashed line) $J$-state 
interference.
Right panel: the $Q/I$ profile calculated in the indicated atmospheric models,
taking into account PRD and $J$-state interference effects.
The reference direction for positive $Q$ is the parallel to the nearest limb.} 
\label{fig:figure-5}
\end{figure} 

\end{document}